\documentclass[11pt]{article}

\usepackage[a4paper,margin=2.5cm]{geometry}

\usepackage{graphicx}
\usepackage{subcaption}
\usepackage[percent]{overpic}

\usepackage{amsmath,amssymb}
\usepackage{natbib}

\usepackage{hyperref}
\usepackage{microtype}   
\usepackage{setspace}    
\usepackage{titlesec}    
\usepackage{authblk}     

\newcommand\BibTeX{{\rmfamily B\kern-.05em \textsc{i\kern-.025em b}\kern-.08em
T\kern-.1667em\lower.7ex\hbox{E}\kern-.125emX}}

\usepackage{moreverb}
\usepackage{gensymb}
\usepackage{graphicx}

\usepackage{subcaption}
\usepackage[percent]{overpic}

\onecolumn
\usepackage[font=small,labelfont=bf]{caption}
\begin{document}

\title{Hybrid weather prediction using spectral nudging toward machine-learning forecasts}

\author{
\textbf{I. Polichtchouk}$^{1}$\thanks{Corresponding author: inna.polichtchouk@ecmwf.int},
\textbf{M. C. A. Clare}$^{2}$,
\textbf{M. Chantry}$^{1}$,
\textbf{E. Gascón}$^{2}$,
\textbf{M. Maier-Gerber}$^{2}$,
\textbf{B. Vanniere}$^{2}$,
\textbf{S. Lang}$^{1}$,
}

\date{$^{1}$European Centre for Medium-Range Weather Forecasts (ECMWF), Reading, UK $^{2}$European Centre for Medium-Range Weather Forecasts (ECMWF), Bonn, Gemany }
\maketitle

\begin{abstract}
 A hybrid approach to numerical weather prediction is investigated, in which the unperturbed physics-based ECMWF Integrated Forecasting System (IFS) is spectrally nudged toward forecasts from a machine-learned weather forecast model, trained to forecast on model levels. Nudging is applied only to the large scales of virtual temperature and vorticity, with the objective of improving large-scale forecast skill while preserving the dynamical and physical behaviour of the underlying physics-based model at smaller scales. Consistent with previous studies, spectral nudging substantially improves large-scale forecast skill relative to the free-running IFS, with gains of up to 1.5 days in the tropics and 12–18 hours in the extra-tropics, and a reduced frequency of forecast busts. These improvements are achieved while preserving forecast variability. The representation of extreme near-surface weather is maintained or improved. Tropical cyclone track forecasts benefit from improved large-scale steering flow, while storm intensity remains comparable to that of the physics-based model and more physically consistent than in pure machine-learned weather forecast models. These results confirm that scale-selective spectral nudging provides a practical pathway for combining machine-learning and physics-based forecasting systems.
\end{abstract}

\section{Introduction} 

Recent machine-learning weather prediction systems have demonstrated substantial improvements in large-scale forecast skill relative to traditional physics-based numerical weather prediction (NWP) models, particularly when evaluated using metrics such as root-mean-square error and anomaly correlation \citep[e.g.,][]{keisler2022forecasting,pathak2022fourcastnet,bi2023accurate,chen2023fuxi,lam2023learning,lang2024aifsA}. However, deterministic (i.e., unperturbed) machine-learning forecasts are often characterised by smoother spatial structure and reduced small-scale variability. This behaviour arises in part from the use of regression-based loss functions such as mean-squared error, which favour solutions that minimise average error and therefore reduce sensitivity to small-scale displacement errors associated with the so-called “double penalty” effect \citep[e.g.,][]{erbert2013, benboullage2024}. The autoregressive rollout used during training to learn multi-step forecasts can further contribute to a progressive loss of small-scale variance \citep{PhilippeMaierGerber}. While this smoothing can improve large-scale verification scores, it can also limit the ability of unperturbed machine-learning models to represent phenomena that depend on sharp gradients or localised diabatic processes, including tropical cyclone intensity and orographic precipitation extremes \citep[e.g.,][]{liu2024evaluation,demaria2024evaluation,pan2025comparative}.

One promising strategy for combining the strengths of machine-learning and physics-based forecasting systems is hybrid prediction, in which machine-learning forecasts provide large-scale guidance while the physics-based model retains responsibility for smaller-scale dynamics and physical processes, and for generating the full set of operational forecast products. \cite{husain2024leveraging} demonstrated such an approach using spectral nudging in the Environment and Climate Change Canada Global Environmental Multiscale (GEM) model, where large-scale fields were relaxed toward forecasts from the machine-learning model GraphCast. In that configuration, GEM’s virtual temperature and horizontal winds in the 850–250 hPa layer were nudged toward GraphCast forecasts for horizontal wavelengths larger than approximately 2250 km. This hybrid configuration substantially improved large-scale forecast skill relative to the free-running GEM model, yielding gains of up to 34 hours in predictive skill while maintaining realistic small-scale structure and improving tropical cyclone track and precipitation forecasts. A similar approach has also been explored using spectral nudging to couple the CMA-GFS model with forecasts from the FuXi machine-learning system \citep{su2026online}.

Using the European Centre for Medium-Range Weather Forecasts (ECMWF) Integrated Forecasting System (IFS) as the physics-based model, and a custom-built version of ECMWF’s Artificial Intelligence Forecasting System (AIFS) \citep{lang2024aifsA} as the machine-learning counterpart, we adopt a hybrid approach similar to that of \cite{husain2024leveraging}. However, the present implementation differs in several respects. The hybridisation is carried out within the native spectral formulation of the IFS, allowing nudging to be applied directly to prognostic spectral coefficients. In addition, we employ a bespoke AIFS-Single model trained on the full set of 137 IFS model levels. This avoids repeated vertical interpolation between pressure levels and model levels and enables nudging to be applied throughout the troposphere, including near the surface. Finally, the vertical extent of nudging is limited dynamically by the diagnosed lapse-rate tropopause rather than by a fixed pressure layer. 

In this study we focus on the deterministic (i.e., unperturbed) forecasting framework and examine how spectral nudging toward machine-learning AIFS-Single forecast modifies the behaviour of the IFS. In particular, we assess its impact on large-scale forecast skill, small-scale structure, forecast busts, and the representation of high-impact weather, including tropical cyclones and surface extremes. We also investigate how spectral nudging interacts with ongoing model development. Although the present study focuses on unperturbed forecasting, the hybrid framework also has important implications for ensemble prediction. In principle, nudging could be applied to a member of the probabilistically trained AIFS-ENS system \citep{lang2024aifsB}; however, the deterministic forecast skill of a single AIFS-ENS member is typically reduced compared to that of AIFS-Single or the unperturbed IFS, and such an approach would therefore not be expected to improve deterministic forecast performance. Extension of the framework to probabilistic forecasting, in which perturbed members are nudged toward AIFS-ENS forecasts, is shown in \cite{polichtchouk2026hybrid}.

This study is structured as follows. Section~\ref{sec:method} describes the IFS and AIFS-Single models, along with the spectral nudging methodology. Section~\ref{sec:case study} presents a brief case study of the extra-tropical wind storm Amy to illustrate the scale-dependent behaviour of the hybrid system. In Section~\ref{sec:large-scale}, we evaluate the forecast skill of the hybrid IFS relative to the operational IFS and AIFS-Single using almost two years of 10-day forecasts. Section~\ref{sec:extremes} presents the impact of nudging on high-impact weather, including tropical cyclone track and intensity, as well as extremes of precipitation, 10-m wind speed, and 2-m temperature. Section~\ref{sec:model dev} examines the interaction between spectral nudging and physics-based model development. Finally, Section~\ref{sec:conclusion} provides summary, discussion and conclusions.

\section{Models and Methods}\label{sec:method}
\subsection{ECMWF IFS}
We use the full-complexity global ECMWF Integrated Forecasting System (IFS), based on cycle 49r1 \citep{Roberts_49r1}, coupled to the 1/4$\degree$ NEMO ocean model. The IFS employs a semi-implicit, semi-Lagrangian formulation \citep{temperton2001two,hortal2002development,diamantakis2022fast} and is horizontally discretised using a spherical harmonic spectral expansion, combined with a cubic-octahedral reduced Gaussian grid \citep{Malardel2016}. All forecasts in this study are run at TCo1279 horizontal resolution, corresponding to a spectral truncation at total wavenumber 1279 and an average grid-spacing of 9~km. Time integration is performed with a time step of $\Delta t$=450s. Vertically, the IFS uses a pressure-based hybrid $\eta$-coordinate with 137 model levels extending from the surface up to 0.01 hPa. The vertical discretization uses a third-order finite element method.

\subsection{AIFS-Single-ML}\label{aifs_single_ml}
ECMWF has developed its own machine-learned global weather forecasting system. Here, AIFS-Single refers to the version of AIFS that is trained via a MSE based loss function. AIFS-Single has been running operationally since February 2024. It uses an encoder-processor-decoder architecture where the encoder maps the input state to a lower-resolution latent space using a Graph Neural Network (GNN) and a Transformer-based processor then forecasts the evolution of this latent state 6 hours forward in time. The decoder projects the updated latent state back to the target resolution. Further architectural details are described in \cite{lang2024aifsA}.

The AIFS-Single model used in this work (hereafter referred to as AIFS-Single-ML) differs from the operational model as it is trained on the full set of 137 IFS model levels instead of 13 pressure levels, thereby avoiding the need for vertical interpolation during nudging. 

Exploratory work indicated that increasing the hidden dimension (from 1024 for the operational AIFS-Single to 1536) led to a substantial improvement in model skill. This increase may be related to the larger number of input and output fields (increased by approximately a factor of 10). This increase in hidden dimension raises the computational cost. However, because nudging is applied only up to total wavenumber 21, the model is trained at O96 resolution (approximately 120 km) rather than the N320 resolution of the operational AIFS-Single (approximately 28 km), resulting in an overall computational cost that remains lower than for operational AIFS-Single.

The AIFS-Single-ML is trained in multiple phases. The model is first pre-trained to forecast 6~h ahead using ERA5 reanalysis data spanning 1979–2022. This is followed by two fine-tuning stages: first on ERA5 reanalysis data spanning 1979–2022, and subsequently on ECMWF operational high-resolution analyses for the period 2016–2023 (see also subsection~\ref{finetuning} for sensitivity to the fine-tuning period). In both fine-tuning stages, the learning rate is kept constant and an auto-regressive rollout approach is used, in which the model is initialised from its own predictions. In this configuration, rollout is performed up to 36h. Experiments with longer rollout horizons (e.g. 72h used in the operational AIFS-Single model) resulted in excessive damping of variability even for total wave numbers smaller than 21, motivating the shorter rollout used here.

\subsection{Vertical coupling strategies}
An important aspect of the hybrid configuration is how machine-learning fields are represented on the IFS vertical grid. Here, we use AIFS-Single-ML, which predicts directly on the 137 IFS model levels, although alternative approaches are also possible.

One option is to use coarse pressure-level output from operational AIFS-Single and interpolate these fields to model levels at each time step, as in \cite{husain2024leveraging}. This introduces additional computational cost (25–30\% when nudged below the tropopause) and restricts nudging to levels above approximately 850~hPa, limiting its impact within the boundary layer. In practice, this leads to reduced short-range forecast skill relative to the model-level formulation, particularly in the first three days, and to regional degradations in tropical stratocumulus regimes.

Another option is to use a dedicated machine-learned vertical interpolator \citep{clare2026vertical} to map pressure-level output to model levels. This avoids the repeated interpolation cost, but residual errors in the boundary-layer thermodynamic structure currently require nudging to be restricted above approximately 900~hPa, limiting improvements in near-surface variables, although no degradation is observed in tropical stratocumulus regimes.

The model-level formulation adopted here avoids vertical interpolation and allows nudging throughout the troposphere, including within the boundary layer. All approaches considered lead to substantial improvements in forecast skill relative to the free running model, but the model-level formulation yields the largest improvements at short lead times, particularly for near-surface variables, while also incurring the lowest computational cost (13\% when nudged below the tropopause). A comparison of these approaches is provided in the Supporting Information.

\subsection{Spectral nudging}
The spectral formulation of the IFS makes it straightforward to apply spectral nudging. In this study, we apply nudging to the prognostic spectral coefficients of virtual temperature ($T_v$) and vorticity ($\zeta$), constraining only the large scales corresponding to total wavenumbers 0–20 (or horizontal wavelengths larger than 2000~km).

Nudging is achieved by adding an extra relaxation tendency on the right-hand side of the spectral prognostic equations for $T_v$  and  $\zeta$, as follows:
\begin{equation}
    \frac{\partial X}{\partial t}=...-\frac{f(t) g(k)}{\tau}(X_{\text{ifs}}-X_{\text{aifs}}),
\end{equation}
where $X \in \{{T_v,\zeta}\}$, $\tau$ is the relaxation time-scale (set to 12 hours, following \cite{husain2024leveraging} and confirmed here as optimal), $f(t)$ is a temporal ramp function, and $g(k)$ is a vertical mask that restricts nudging below the tropopause:
\begin{equation}
    g(k)=\frac{1}{1+\exp(k_{\text{trp}}+5-k)},
\end{equation}
where $k = 1,\dots,137$ is the model level index, and $k_{\text{trp}}$ denotes the level of the lapse-rate tropopause, which varies in space and time. The temporal ramp function $f(t)$ is:
\begin{equation}
    f(t)=\frac{1}{2}\left( \tanh\left[\frac{t} {t_{\text{ramp}}} - 1 \right ] + 1.0 \right),
\end{equation}
where $t$ is continuous forecast time, with $t=j \Delta t$ in the discrete formulation; $\Delta t = 450 \, \text{s}$ is the IFS time step size, and $t_{\text{ramp}}=8640$~s. This time-ramp gradually increases the nudging strength over the first 8–12 hours of the forecast, reaching full strength around 12 hours. This approach mitigates degradation against ECMWF analysis in early forecast skill, as the free-running IFS typically performs better than the nudged configuration immediately after initialisation.

The nudging is applied at every model time step. Since AIFS-Single predictions are only available at 6-hourly intervals, linear temporal interpolation is used to compute  $X_\text{aifs}$  at intermediate times. In practice, a 10-day AIFS forecast is first run, initialised from the operational ECMWF analysis, with output saved every 6 hours for specific humidity ($q$), temperature ($T$), and horizontal wind components $(u, v)$ on all 137 model levels. The virtual temperature is then computed in grid-point space as  $T_v = T (1 + 0.61q)$ . This  $T_v$  field, along with  $(u,v)$, is transformed into spectral space to obtain the spectral coefficients of $T_v$ and $\zeta$, which are used as the nudging target fields.

Once the nudging input fields are prepared, an IFS forecast is run with spectral nudging activated. We found that nudging the zonal wavenumber  $m = 2$ of divergence introduced aliasing of the semi-diurnal tide signal. This resulted in a wavenumber-2 pattern of degradation in tropical forecasts of mean sea level pressure and geopotential height. To avoid this artifact, divergence is excluded entirely from the nudging procedure, as experiments in which divergence was nudged while excluding $m=2$ did not lead to any additional improvement in forecast performance.

\subsection{Experimental setup}
Nudged and un-nudged 10-day forecasts are initialised every day for over one and a half years between 1 June 2024 and 28 February 2026 at 00 UTC. Standard forecast skill score metrics are computed by the in-house verification software QUAVER. Scores against ECMWF analysis are computed on a 1.5$\degree \times$ 1.5$\degree$ grid. For verification against radiosonde and SYNOP observations, full-resolution (O1280) model fields are used, with model values interpolated to observation locations using nearest-neighbour interpolation. Tropical cyclones are tracked by the method described in \cite{vitart2012new} and \cite{magnusson2021tropical}.

\begin{figure}
\centering
\includegraphics[width=13cm]{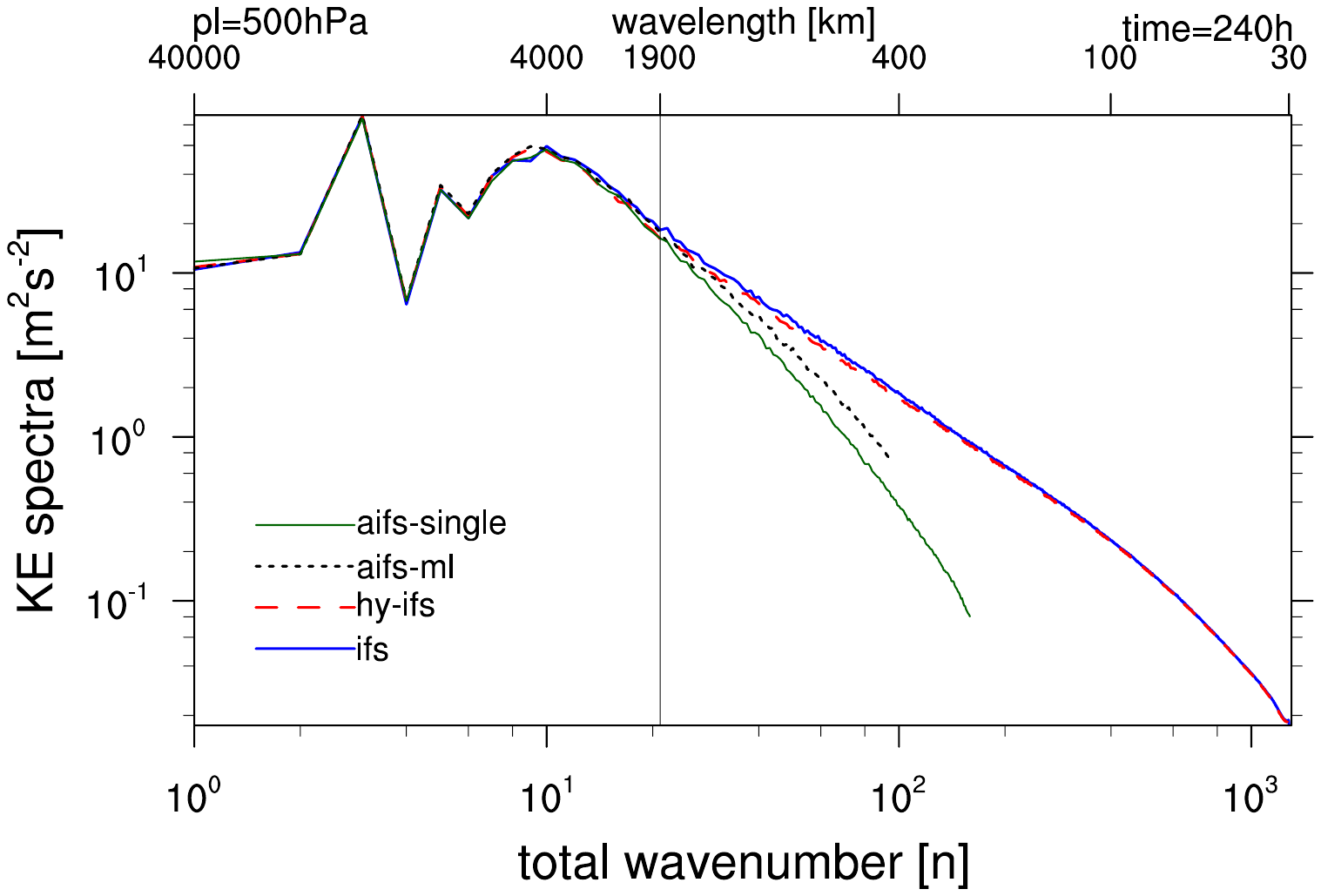}
\caption{Global kinetic energy spectra as a function of total wavenumber at 500~hPa for forecast day 10 from AIFS-Single-ML (dotted black), IFS (blue), hy-IFS (dashed red) and operational AIFS-Single (dark green). Spectra are averaged for 3 months of forecasts. Black vertical line shows total wavenumber 21, which is the cut-off wavenumber for nudging.}\label{fig:fig1}
\end{figure}

To illustrate that the nudging approach does not substantially affect the representation of small scales, kinetic energy wavenumber spectra for the IFS, the nudged IFS (hy-IFS), the bespoke AIFS-Single-ML used for nudging, and the operational AIFS-Single are shown in Figure~\ref{fig:fig1} for day 10 of the forecast. The hy-IFS closely follows the IFS kinetic energy spectrum, indicating that the nudging does not introduce additional damping at small scales. In contrast, the operational AIFS-Single exhibits reduced energy beyond total wavenumbers of about 20. The AIFS-Single-ML used for nudging retains slightly more energy at these scales than the operational AIFS-Single, which is likely related to its shorter rollout length (36 hours vs. 72 hours).

\subsection{Computational cost}
The application of spectral nudging increases the computational cost of the IFS integration by approximately 13\%. This increase arises primarily because the nudging increments are computed in spectral space but must be applied below a diagnosed lapse-rate tropopause that varies with latitude and longitude. The increment therefore has to be transformed to grid-point space at each model time step so that the tropopause-dependent vertical mask can be applied. This additional spectral-to-grid-point transform constitutes the dominant computational overhead.  The cost associated with generating the AIFS-Single-ML forecasts used as the nudging reference is comparatively small.

An alternative configuration would be to apply nudging below a fixed model level, avoiding the need for a latitude- and longitude-dependent mask and reducing the computational overhead to approximately 2.5\%. However, because the tropopause height varies substantially with latitude, such an approach would require nudging to be limited to levels below approximately 300 hPa in order to avoid degrading the stratospheric circulation where the AIFS-Single model performs suboptimally. This would substantially reduce the nudging influence in the tropical upper troposphere, where some of the largest forecast skill improvements are obtained. The configuration adopted here therefore represents a compromise between computational efficiency and forecast performance.

\section{Case study: Extra-tropical cyclone Amy}\label{sec:case study}
Before presenting the statistical evaluation, we illustrate the differences between IFS nudged to AIFS-Single-ML (i.e., hy-IFS), IFS, and operational AIFS-Single using a representative case study, complementing the kinetic energy spectra analysis in Figure~\ref{fig:fig1}. Storm Amy was the first north-western European wind storm of the 2025 season, affecting Ireland, Scotland and Norway on 3–4 October 2025 with severe winds. The cyclone originated from a synoptic-scale disturbance associated with the remnants of tropical cyclones Humberto and Imelda over the western Atlantic. It propagated rapidly eastward and intensified on 3 October, reaching northern Ireland and western Scotland later that day.

\begin{figure}
\centering
\includegraphics[width=0.8\textwidth]{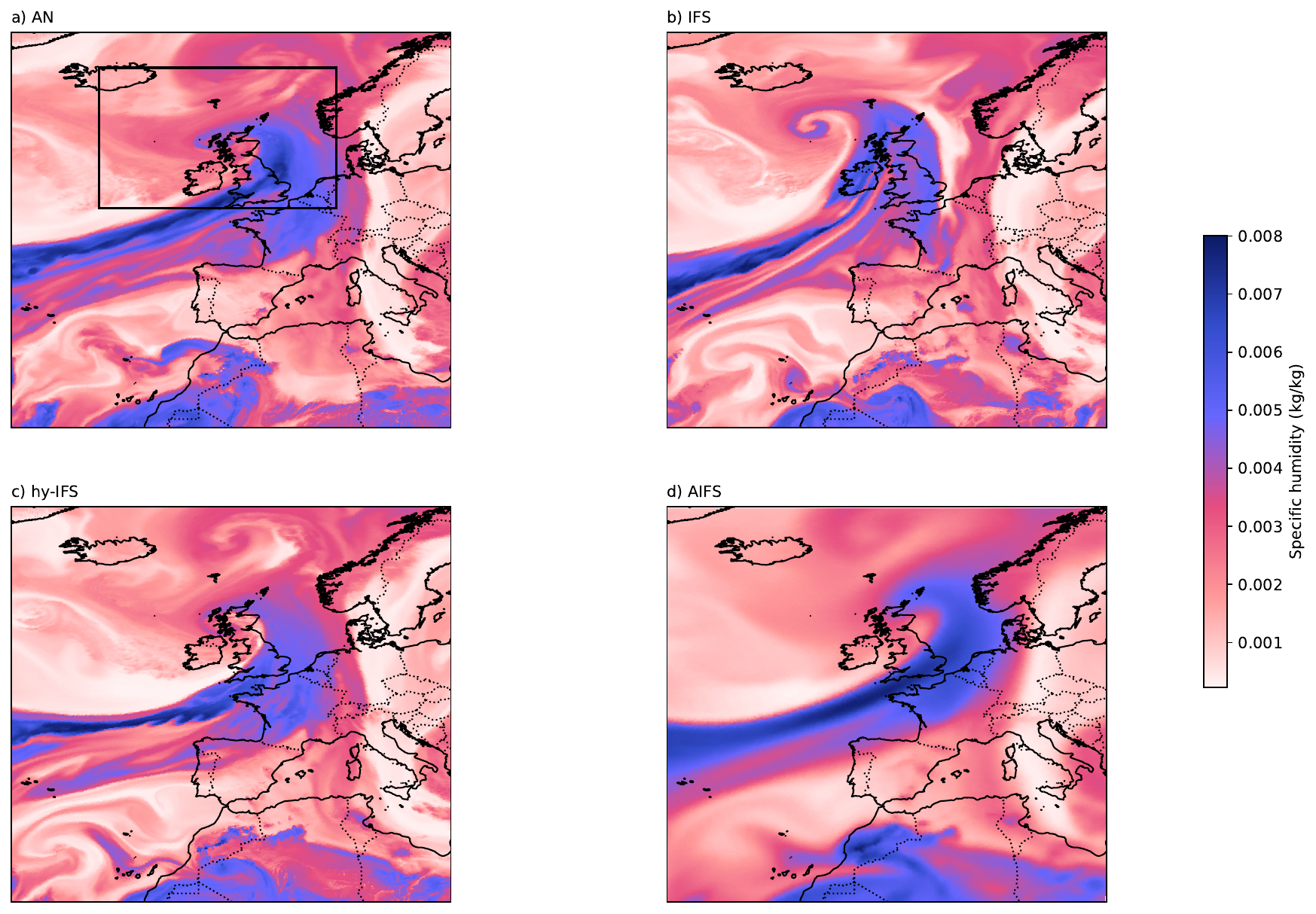}
\caption{Specific humidity at 700~hPa over western Europe at 18 UTC on 3 October 2025 in (a) the ECMWF analysis and t=114~h forecasts by (b) IFS, (c) hy-IFS, and (d) AIFS. The black box in (a) marks the region shown in Figure~\ref{fig:fig3}. The hy-IFS retains sharper gradients and mesoscale structure than the operational AIFS-Single forecast.}
\label{fig:fig2}
\end{figure}

\begin{figure}[ht!]
\centering
\includegraphics[width=0.8\textwidth]{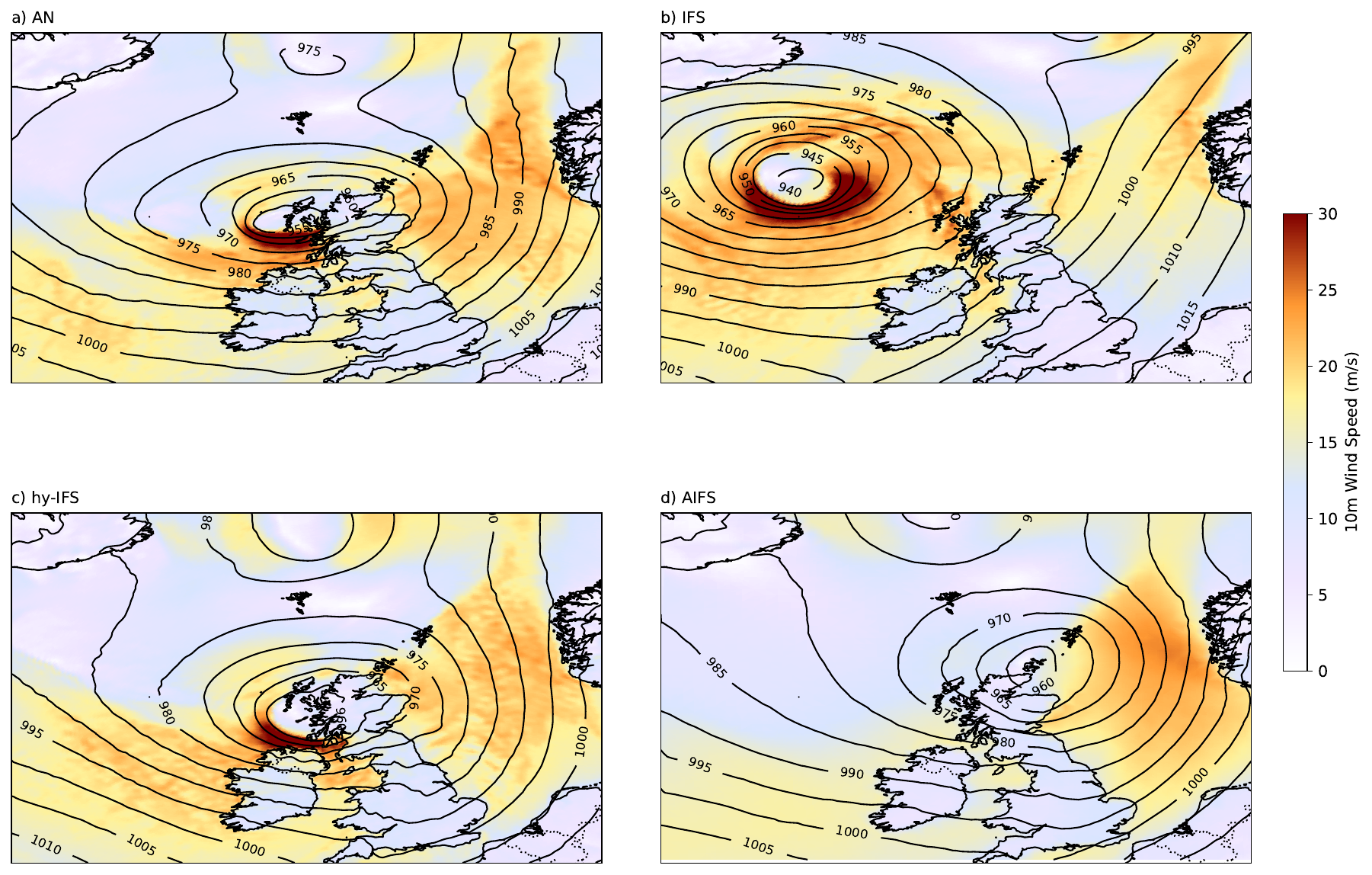}
\caption{Mean sea level pressure (black contours) and 10~m wind speed (shading) over the region marked by black box in Figure~\ref{fig:fig2}a for (a) ECMWF analysis and t=114~h forecasts by (b) IFS, (c) hy-IFS, and (d) AIFS. The pure AIFS forecast underestimates the near-surface wind maxima over western Scotland and out of the three forecasts, the hy-IFS most closely reproduces the analysed storm structure and near-surface wind maxima. }
\label{fig:fig3}
\end{figure}

The 700 hPa specific humidity and 10-m wind speed fields associated with Storm Amy at 18 UTC on 3 October are shown for the ECMWF analysis in Figures \ref{fig:fig2}–\ref{fig:fig3}a, with the corresponding 114 h forecasts from the three models shown in Figures \ref{fig:fig2}–\ref{fig:fig3}c–d. The AIFS-Single forecast is markedly smoother and lacks the small-scale structure present in the analysis, consistent with the kinetic energy spectra in Figure \ref{fig:fig1}. By contrast, the hy-IFS retains sharp gradients and mesoscale features similar to IFS, while exhibiting a large-scale structure closer to the analysis, as also seen in the mean sea level pressure field (Figure \ref{fig:fig3}). The AIFS-Single forecast furthermore underestimates the near-surface wind maxima over western Scotland, unlike both IFS and hy-IFS. For this case, the hy-IFS provides the closest overall agreement with the storm structure in the analysis.

This example highlights the ability of the hy-IFS to combine improved large-scale evolution of machine-learning models with realistic mesoscale structure of physics-based models. We now examine whether this behaviour is reflected in the statistical verification of large-scale forecast skill and in the simulation of extreme weather events.


\section{Large-scale skill}\label{sec:large-scale}
\begin{figure}
\centering
\includegraphics[width=0.7\textwidth]{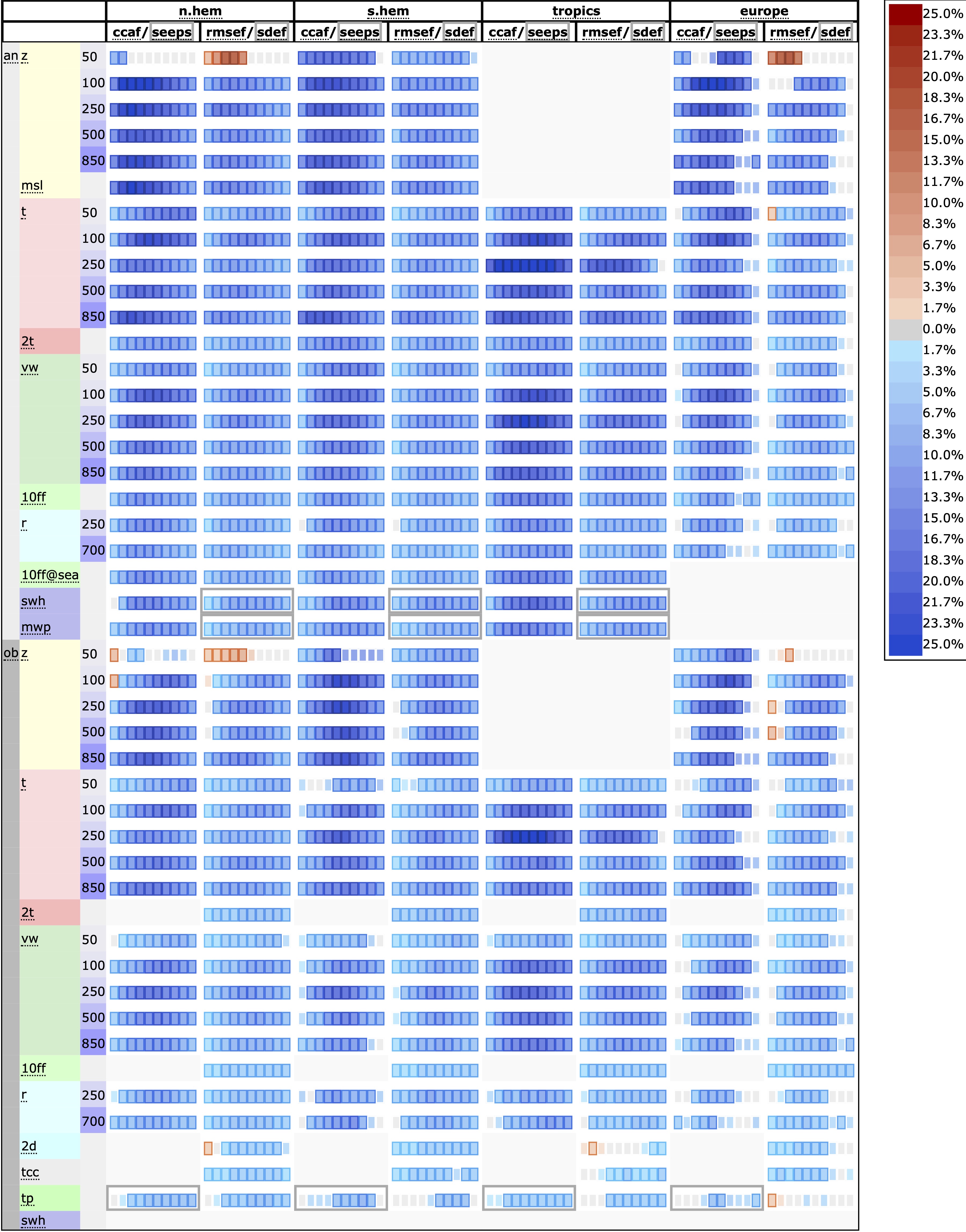}
\caption{Summary scorecard comparing forecast skill differences between the spectrally nudged IFS (hy-IFS) and the control IFS, using anomaly correlation (CCAF) and root mean square error (RMSEF), verified against the ECMWF operational analysis (top) and against radiosonde and SYNOP observations (bottom). Shades of blue indicate improvement in hy-IFS relative to IFS, and shades of red indicate degradation. Each box corresponds to a 24-h lead-time interval from forecast day 1 to day 10. Colour saturation reflects the magnitude of the normalized difference. Statistically significant differences at the 99.7\% confidence level are outlined with coloured frames.  Regional abbreviations denote Northern Hemisphere (n.hem; 20$\degree$N-90$\degree$N), Southern Hemisphere (s.hem; 20$\degree$S–90$\degree$S), Tropics (20$\degree$N-20$\degree$S) and Europe (35$\degree$N-75$\degree$N, 12.5$\degree$W-42.5$\degree$E). Scores are computed from over 644 forecasts initialized at 00 UTC between 1 June 2024 and 28 February 2026. }
\label{fig:fig4}
\end{figure}

\begin{figure}
\centering
\includegraphics[width=\textwidth]{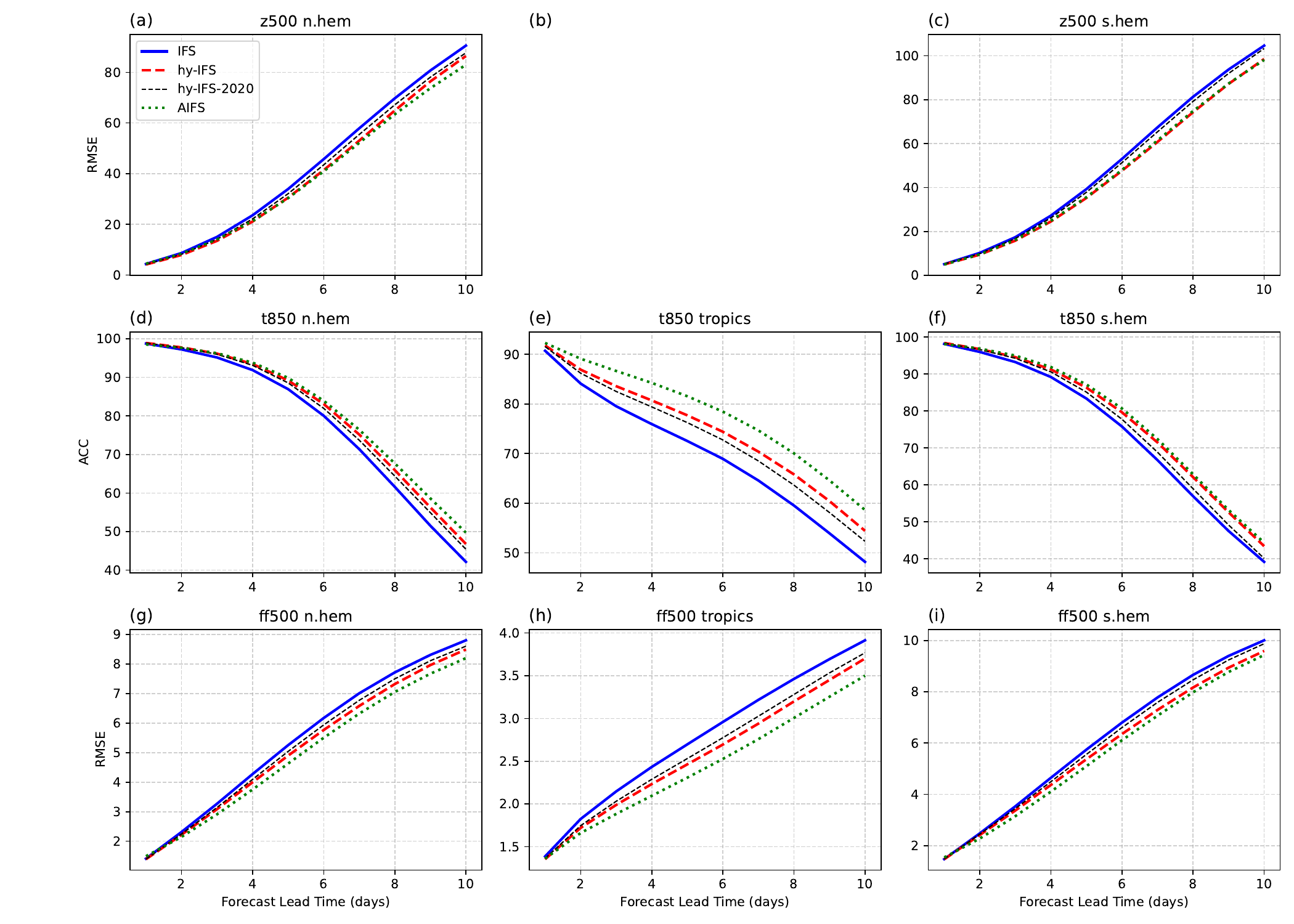}
\caption{RMSE of 500~hPa geopotential height (top); ACC of 850~hPa temperature (middle); and RMSE of 500~hPa wind speed (bottom) for hy-IFS (dashed red), IFS (solid blue) and operational AIFS (dotted green) for the (a,d,g) Northern Hemisphere extra-tropics (20$\degree$N-90$\degree$N), (b,e,h) tropics (20$\degree$N-20$\degree$S), and (c,f,i) Southern Hemisphere extra-tropics (20$\degree$S-90$\degree$S). Thin black lines show hybrid forecasts with AIFS-Single-ML fine-tuned on ECMWF analysis only for 2016–2020, rather than 2016–2023 as in hy-IFS (dashed red). Lower RMSE values and higher ACC values indicate improved forecast accuracy. Scores are computed against ECMWF operational analysis for forecasts initialized at 00 UTC between 1 June 2024 and 28 February 2026. }
\label{fig:fig5}
\end{figure}

\begin{figure}
\centering
\includegraphics[width=\textwidth]{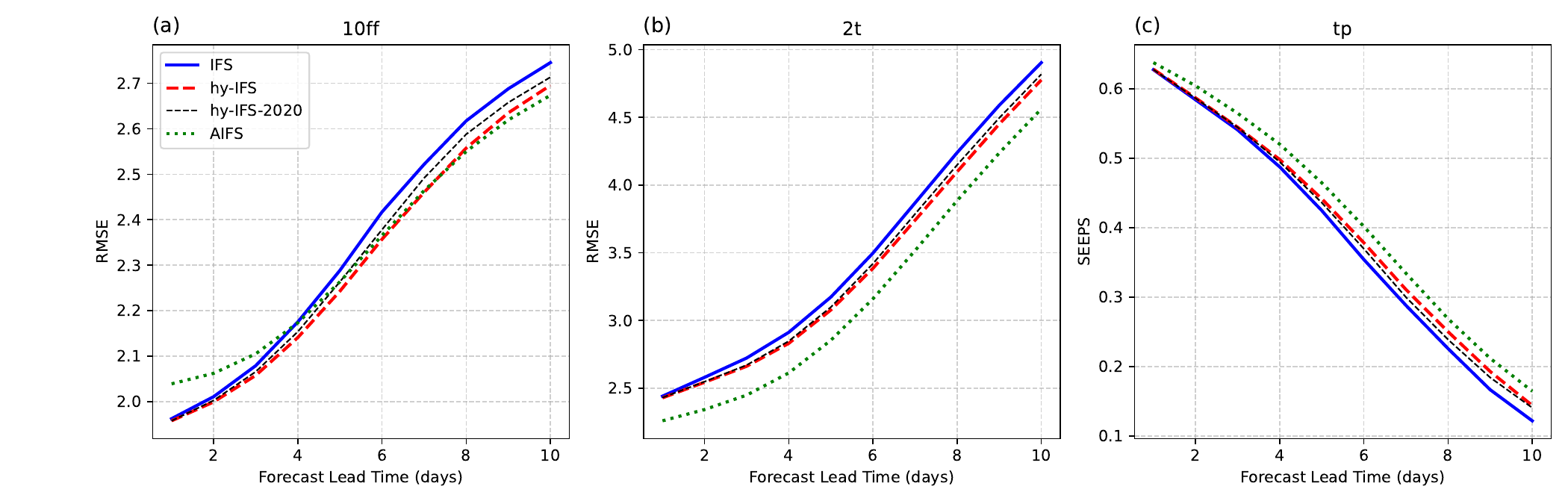}
\caption{RMSE of (a) 10-m wind speed and (b) 2-m temperature, and SEEPS of (c) total precipitation for hy-IFS (dashed red), IFS (solid blue) and operational AIFS (dotted green) for the Northern Hemisphere extra-tropics. Thin black lines show hybrid forecasts with AIFS-Single-ML fine-tuned on ECMWF analysis only for 2016–2020, rather than 2016–2023 as in hy-IFS (dashed red).  Lower RMSE values and higher SEEPS values indicate improved forecast accuracy. Scores are computed against SYNOP station observations from over 644 forecasts initialized at 00 UTC between 1 June 2024 and 28 February 2026. }
\label{fig:fig6}
\end{figure}
Figure~\ref{fig:fig4} presents a summary scorecard for selected variables averaged over several regions, comparing the hy-IFS against IFS. Verification is shown against the ECMWF operational analysis (upper half) and against radiosonde and SYNOP station observations (lower half). Forecast skill is evaluated using root-mean-squared error (RMSE) and anomaly correlation (ACC).

Nudging the IFS toward AIFS-Single-ML results in a predominantly blue scorecard, indicating widespread improvements in forecast skill. The largest gains occur in the tropics, where upper-air scores for temperature and wind improve by up to 33\% in ACC and by up to 20\% in RMSE, particularly in the 850–250 hPa layer. Seasonal analysis (not shown) indicates that outside the tropics the largest improvements occur in the summer hemisphere. Improvements in surface parameters, including 2-m temperature and 10-m wind speed (especially over ocean), reach up to 10\% for ACC and 8\% for RMSE when verified against the ECMWF operational analysis. When verified against SYNOP station observations, improvements are likewise positive but smaller in relative terms, typically in the range of 3–5\%.

Although spectral nudging is applied only below the tropopause, the hybrid system also exhibits improved forecast skill in the lower stratosphere. Improvements are evident across most fields at 100 hPa and 50 hPa, with the exception of geopotential height RMSE at 50 hPa, which shows a degradation associated with a small shift in mean bias. This behaviour is consistent with the fact that improved tropospheric evolution provides more accurate lower-boundary forcing for the stratosphere, and similar behaviour is found in the hybrid ensemble configuration \citep{polichtchouk2026hybrid}.

To further quantify these improvements and to compare hy-IFS with the operational machine-learning model AIFS-Single, Figure~\ref{fig:fig5} shows selected upper-air scores as a function of lead time for the models in the Northern Hemisphere extra-tropics, tropics, and Southern Hemisphere extra-tropics. Relative to IFS, hy-IFS exhibits forecast skill gains of up to approximately 1.5 days in the tropics and up to 12-18 hours in the extra-tropics in both hemispheres. In operational physics-based NWP, improvements of this magnitude are substantial and are comparable to gains typically achieved over many years of model development \citep[e.g.,][]{bauer2015quiet}. 

Hybrid IFS shows broadly comparable skill in the extra-tropics to AIFS-Single, where AIFS-Single exhibits a marginal advantage. In the tropics, AIFS-Single retains a clearer advantage over hy-IFS, corresponding to roughly one additional day of forecast skill. Since spectral nudging is applied for wavelengths larger than approximately 2000~km (see Figure~\ref{fig:fig1}), the superior tropical performance of AIFS-Single indicates that it retains additional skill at smaller scales that are not constrained in hy-IFS. A similar behaviour is seen in the ensemble framework, where AIFS-ENS outperforms its hybrid counterpart in the tropics \citep{polichtchouk2026hybrid}, despite not exhibiting comparable smoothing to AIFS-Single for wavelengths smaller than 2000~km, indicating that the differences are not due to smoothing alone. We note that no scale-selective verification has been applied here, in contrast to \cite{husain2024leveraging}, where only scales with comparable spectral energy in physics-based, hybrid, and machine-learning models are evaluated.

Figure~\ref{fig:fig6} shows the evolution of forecast error metrics with lead time for selected surface variables in the Northern Hemisphere extra-tropics for the models. Relative to IFS, hy-IFS exhibits substantial improvements in RMSE of 10-m wind speed and 2-m temperature, corresponding to an approximate 6–12 h gain in forecast skill. Improvements are also evident for total precipitation as measured by Stable
and Equitable Error in Probability Space (SEEPS) \citep{rodwell2010}. Total precipitation is evaluated using SEEPS rather than RMSE because precipitation is highly intermittent and non-Gaussian, and RMSE can artificially favour smoother forecasts through reduced amplitude variability, whereas SEEPS provides a more appropriate assessment of precipitation occurrence and intensity categories. The hy-IFS outperforms AIFS-Single for 10-m wind speed up to day 5, while AIFS-Single retains a clear advantage for 2-m temperature and total precipitation throughout most of the forecast range. The weaker performance of AIFS-Single for 10-m wind speed is likely linked to its lower horizontal resolution (28 km versus 9 km) and to the fact that 10-m wind observations are not directly assimilated in either ERA5 or the ECMWF operational analysis used for training and initialization. A similar result was also found for the hybrid ensemble system in \cite{polichtchouk2026hybrid}.

We note that these surface variables are not directly nudged in hy-IFS and are therefore only indirectly constrained through the large-scale temperature and wind fields on model levels. In contrast, machine-learning forecasts predict near-surface variables directly from (re)analyses, which likely contributes to their stronger performance for thermodynamic surface quantities such as 2-m temperature and precipitation.

\subsection{Sensitivity to the AIFS training period}\label{finetuning}
The performance of the hybrid system depends on the training of the machine-learning model used for nudging. In the results above, AIFS-Single-ML is fine-tuned on ECMWF analysis data from 2016–2023. To assess sensitivity to the fine-tuning period, we repeat the experiments using a model fine-tuned only on 2016–2020. The impact is also shown in Figures~\ref{fig:fig5}–\ref{fig:fig6}, where thin black lines denote the reduced training-period configuration and red dashed lines the reference configuration.

Including more recent training data leads to systematic improvements in forecast skill, reaching up to 10\%, particularly in the tropics and Southern Hemisphere, indicating sensitivity to the training data.

These results suggest that the machine-learning model used for nudging should be periodically retrained to incorporate more recent data. The relative roles of increased sample size and improved representativeness cannot be disentangled here. In particular, significant changes in the operational analysis may require retraining to maintain consistency between the machine-learning model and the analysis used for initialization. However, even with the reduced training period, the hybrid system consistently outperforms the un-nudged IFS.

\subsection{Forecast busts}\label{sec:fc busts}
\begin{figure}
\centering
\includegraphics[width=0.9\textwidth]{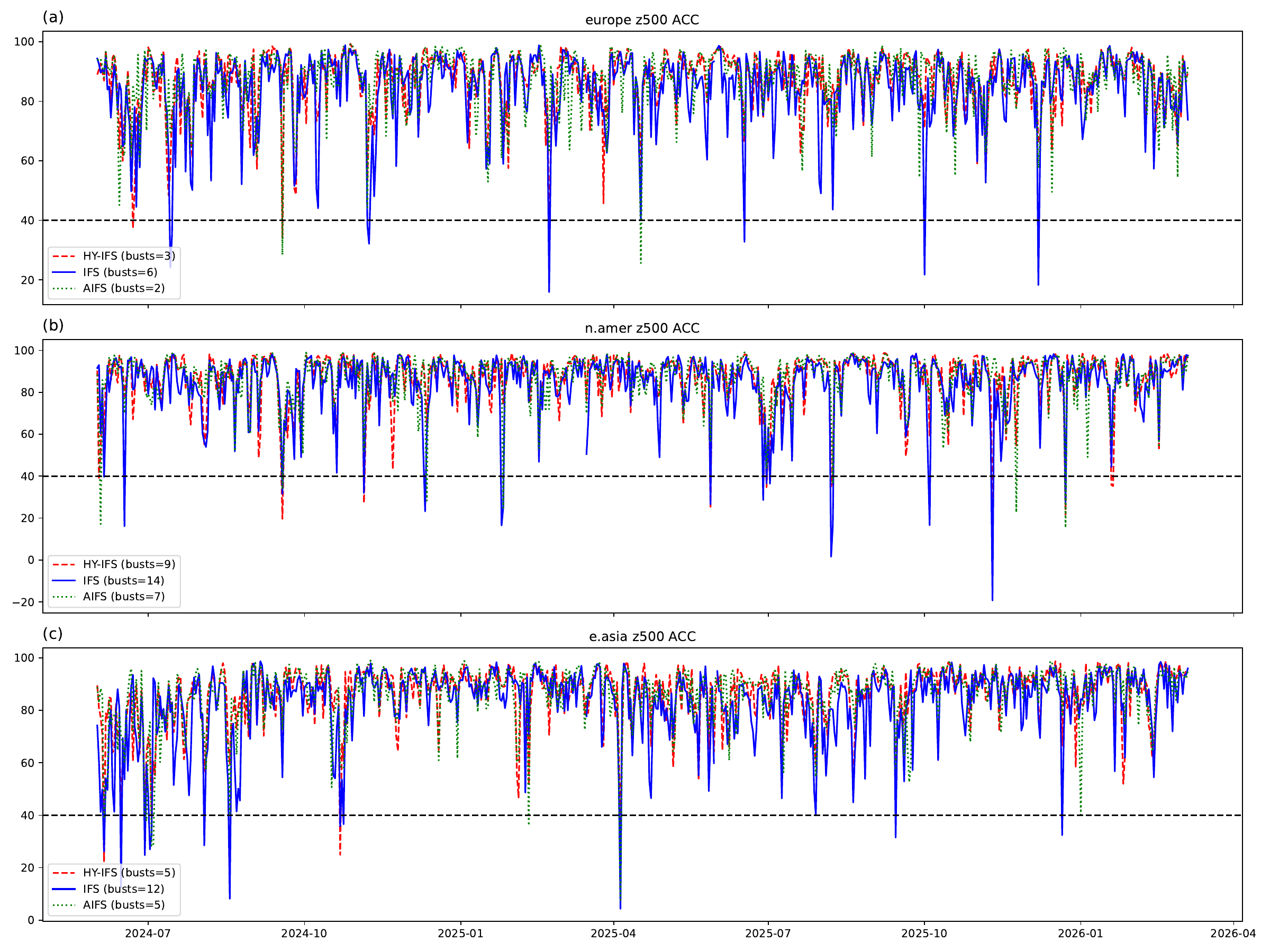}
\caption{Timeseries of day 6 ACC of geopotential height for hy-IFS (red), IFS (blue) and operational AIFS (green) for (a) Europe (35$\degree$N-75$\degree$N; 12.5$\degree$W-42.5$\degree$E), (b) North America (25$\degree$N-60$\degree$N; 120$\degree$W-75$\degree$W) and (c) East Asia (25$\degree$N-60$\degree$N; 102.5$\degree$E-150$\degree$E). A forecast bust is defined to occur when ACC drops below 40\%. The number of busts for each system is shown in the legend.  }
\label{fig:fig7}
\end{figure}

While mean skill scores provide a useful measure of average forecast performance, they do not focus on rare but high-impact forecast busts. We therefore examine forecast busts to assess whether spectral nudging improves the robustness of large-scale flow prediction. Forecast busts are defined here following \cite{rodwell2013characteristics} as cases where the anomaly correlation coefficient (ACC) of 500 hPa geopotential height falls below 40\% at forecast day 6. Figure~\ref{fig:fig7} shows the occurrence of forecast busts over Europe, North America, and East Asia for forecasts initialised between 1 June 2024 and 28 February 2026 for IFS, hy-IFS, and operational AIFS-Single.

Across all regions, the number of forecast busts is substantially reduced in hy-IFS compared to IFS. The number of events decreases from six to three over Europe, from fourteen to nine over North America, and from twelve to five over East Asia, corresponding to an approximate factor-of-two reduction in bust frequency in hy-IFS. The number of busts in hy-IFS and operational AIFS-Single is broadly comparable, with AIFS-Single showing a marginally smaller number of events.

Forecast busts in hy-IFS and operational AIFS-Single do not necessarily occur on the same dates. This reflects differences between the AIFS-Single-ML model used for nudging and the operational AIFS-Single model, leading to distinct forecast evolution and errors, despite similar overall bust frequencies.

\section{Results: Extreme event prediction}\label{sec:extremes}
\subsection{Tropical Cyclones}
\begin{figure}
\centering
\includegraphics[width=\textwidth]{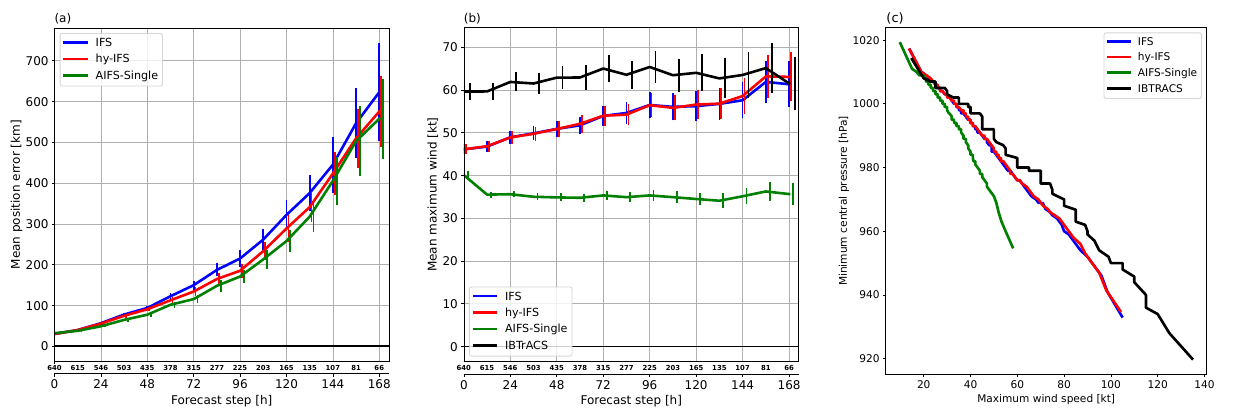}
\caption{(a) Mean position error and (b) mean maximum wind for tropical cyclones as a function of forecast lead time for the hy-IFS (in red), IFS (in blue), and AIFS-Single (in green). Verification statistics are computed with respect to IBTrACS \citep{Gahtan,knapp2010international} for forecasts initialised at 00 UTC between 1 June 2024 and 31 October 2025. Case counts for each lead time are displayed below the graphs and vertical bars show 2.5\%-97.5\% confidence intervals. (c) Minimum central pressure versus maximum wind speed relationship for tropical cyclones as a function of forecast lead time.}
\label{fig:fig8}
\end{figure}

While improvements in large-scale forecast skill and reductions in forecast bust frequency indicate improved large-scale performance of the hybrid system, it is of interest to assess whether these gains are associated with improved prediction of high-impact weather. Tropical cyclones provide a stringent test in this regard, as accurate forecasts require both realistic large-scale steering flow and the preservation of mesoscale structure. In this section, we evaluate the impact of spectral nudging on tropical cyclone track and intensity forecasts and compare the performance of hy-IFS with IFS and operational AIFS-Single (Figure~\ref{fig:fig8}).

Consistent with previous studies with other models \citep{husain2024leveraging,liu2024hybrid,niu2025machine,lai2025towards,su2026online}, spectral nudging substantially improves tropical cyclone track forecasts relative to IFS, while leaving intensity forecasts largely unchanged. Track errors are reduced by up to approximately 50 km, corresponding to an improvement of roughly 12 hours in forecast lead time (Figure~\ref{fig:fig8}a). This behaviour is consistent with the improved large-scale skill discussed in Section~\ref{sec:large-scale}, which leads to a more accurate representation of the steering flow. Operational AIFS-Single exhibits slightly better track forecasts than hy-IFS, possibly due to its higher tropical large-scale skill (Figure~\ref{fig:fig5}).

The advantage of the hybrid system lies in maintaining physically consistent intensity evolution. While track prediction improves relative to IFS, the relationship between minimum central pressure and maximum wind speed remains comparable to that of the physics-based model, whereas the pure machine-learning forecasts tend to under-predict storm intensity despite improved track accuracy (Figure~\ref{fig:fig8}c).

\subsection{Northern Hemisphere surface variables}

\begin{figure}
\centering
\includegraphics[width=\textwidth]{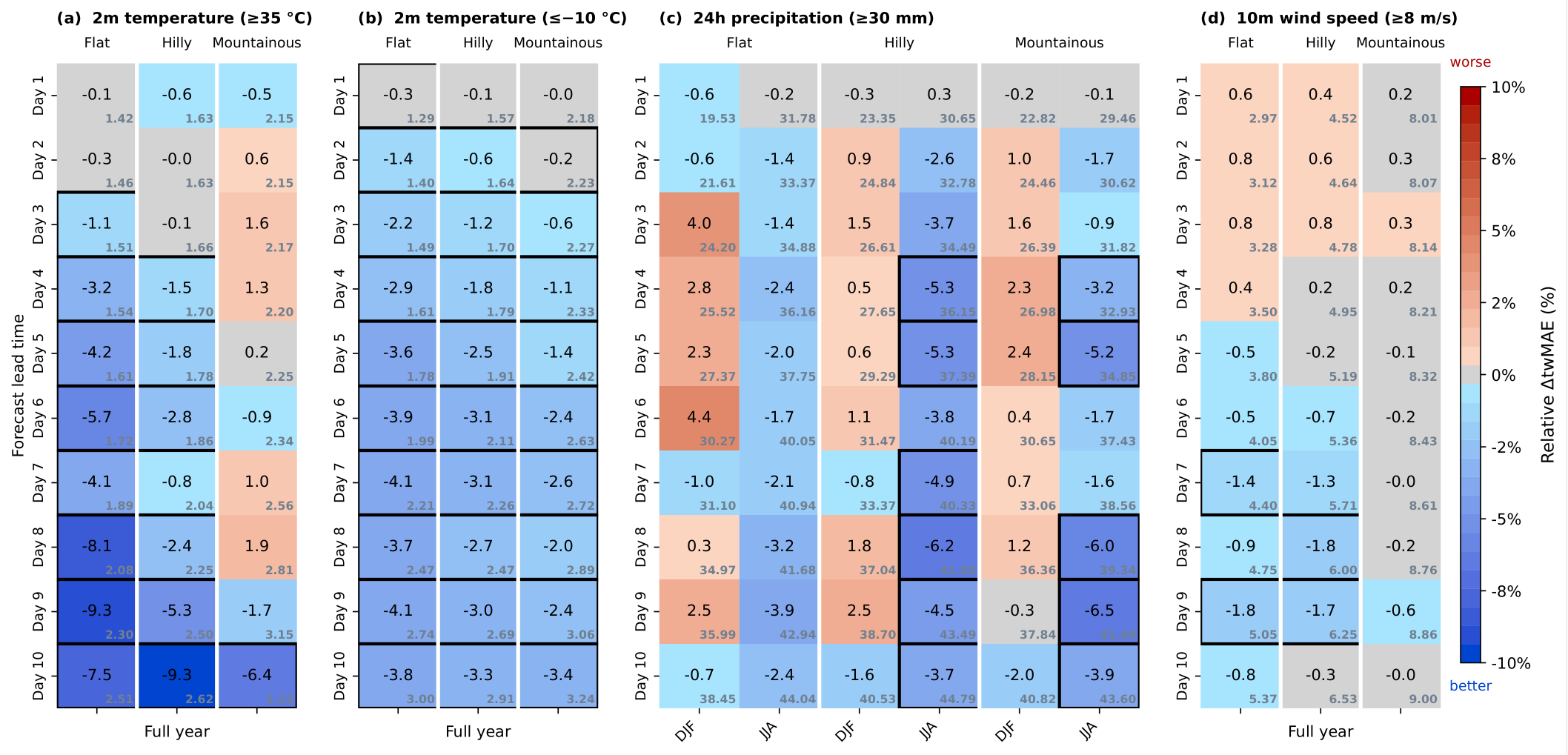}
\caption{Relative change (\%) in threshold-weighted mean absolute error (twMAE) for extreme near-surface variables in the Northern Hemisphere, expressed as percentage difference (hy-IFS minus IFS) relative to IFS. Results are shown as a function of forecast lead time (days 1–10) and stratified by orographic complexity (flat, hilly, mountainous). Panels show (a) extreme high 2-m temperature events ($\geq35\degree$C), (b) extreme low 2-m temperature events ($\leq-10\degree$C), (c) heavy precipitation events ($\geq30$~mm/day), and (d) strong 10-m wind events ($\geq8$ m/s). Blue (red) shading indicates improved (degraded) skill of the hybrid system relative to IFS. Numerical values denote relative differences in twMAE (\%), with the corresponding twMAE hy-IFS raw values indicated in smaller text at the bottom right. Statistically significant differences are highlighted with bold black frames. Precipitation results are separated by season to account for differing precipitation regimes, while wind speed and 2-m temperature results are aggregated over all seasons to increase sample size. Verification is against SYNOP station observations. }
\label{fig:fig9}
\end{figure}

Figure~\ref{fig:fig9} summarises the impact of spectral nudging on the representation of extreme near-surface weather events in the Northern Hemisphere, evaluated against SYNOP station observations. Forecast errors are quantified using the threshold-weighted Mean Absolute Error (twMAE), a consistent scoring rule for deterministic forecasts that emphasises values exceeding a prescribed threshold~\citep{Ehm2016, Allen2023}. The relative change (\%) in twMAE for extreme near-surface variables (hy-IFS minus IFS, relative to IFS) is shown in the figure.

Fixed absolute thresholds were selected for each variable to target rare events. Thresholds of 35\,°C and $-10$\,°C are used for warm and cold 2-m temperature extremes, respectively~\citep{Sherwood2010, Zhang2011, Lehmann2025, WMO2020extremes}. For 10-m wind speed, a threshold of 8\,m\,s$^{-1}$ corresponds to the 99th percentile of the observed climatology. For precipitation, a daily threshold of 30 mm is consistent with the WMO/ETCCDI index Rnnmm~\citep{Zhang2011}. Orographic complexity is characterised by the standard deviation of filtered subgrid orography (\textsc{SDfor}), derived from 1\,km elevation data by filtering out features larger than 5\,km. Terrain types are defined as flat (\textsc{SDfor}\,$<$\,40\,m), hilly (40–120\,m), and mountainous (\textsc{SDfor}\,$>$\,120\,m).

Overall, the hybrid configuration shows consistent improvement for temperature extremes, with reduced errors for both warm and cold events across most lead times and orographic regimes, with the largest gains over flat terrain, reaching up to 9\%. Improvements grow with lead time, suggesting that the machine-learning large-scale guidance is associated with more accurate near-surface temperature evolution at medium range.

For heavy precipitation, the impact of nudging is more mixed. Improvements of up to 5\% are found over hilly and mountainous regions in summer (JJA), while minor, statistically non-significant degradations are observed in winter (DJF). For strong near-surface wind, hy-IFS shows a small but generally consistent improvement at longer lead times, predominantly over flat and hilly terrain.

Although at least 1000 observation samples exceeding the thresholds were considered for each evaluated condition, these results should be interpreted with caution, as the verification of extreme events in deterministic forecasts is inherently affected by large sampling uncertainty due to their rarity, and traditional performance measures can degrade as events become rarer~\citep{Ferro2011}. A more robust assessment would ideally be conducted within a probabilistic ensemble framework, where the full forecast uncertainty can be properly characterised~\citep{BenBouallegueMASS2024}.

\section{Impact of spectral nudging on physics-based model development}\label{sec:model dev}
A potential concern when introducing spectral nudging into a forecasting system is that it may complicate the evaluation of physics-based model development. Because nudging constrains the large-scale circulation toward an external forecast, it could in principle mask degradations introduced by changes in physical parametrisations, thereby making it more difficult to diagnose the impact of model modifications using standard forecast verification metrics. This raises the question of whether physics development becomes more difficult in a hybrid configuration.

To investigate this issue, we consider a controlled physics change in which the entrainment rate in the convection scheme is doubled. This modification improves aspects of small-scale convective organization but leads to a degradation in large-scale forecast skill, particularly in the tropics, when evaluated in the standard IFS configuration without nudging. Since this change affects not only convective-scale behaviour but also the large-scale circulation through modifications to diabatic heating and moisture transport, it provides a useful test of whether spectral nudging suppresses or obscures such signals.

\begin{figure}
\centering
\includegraphics[width=\textwidth]{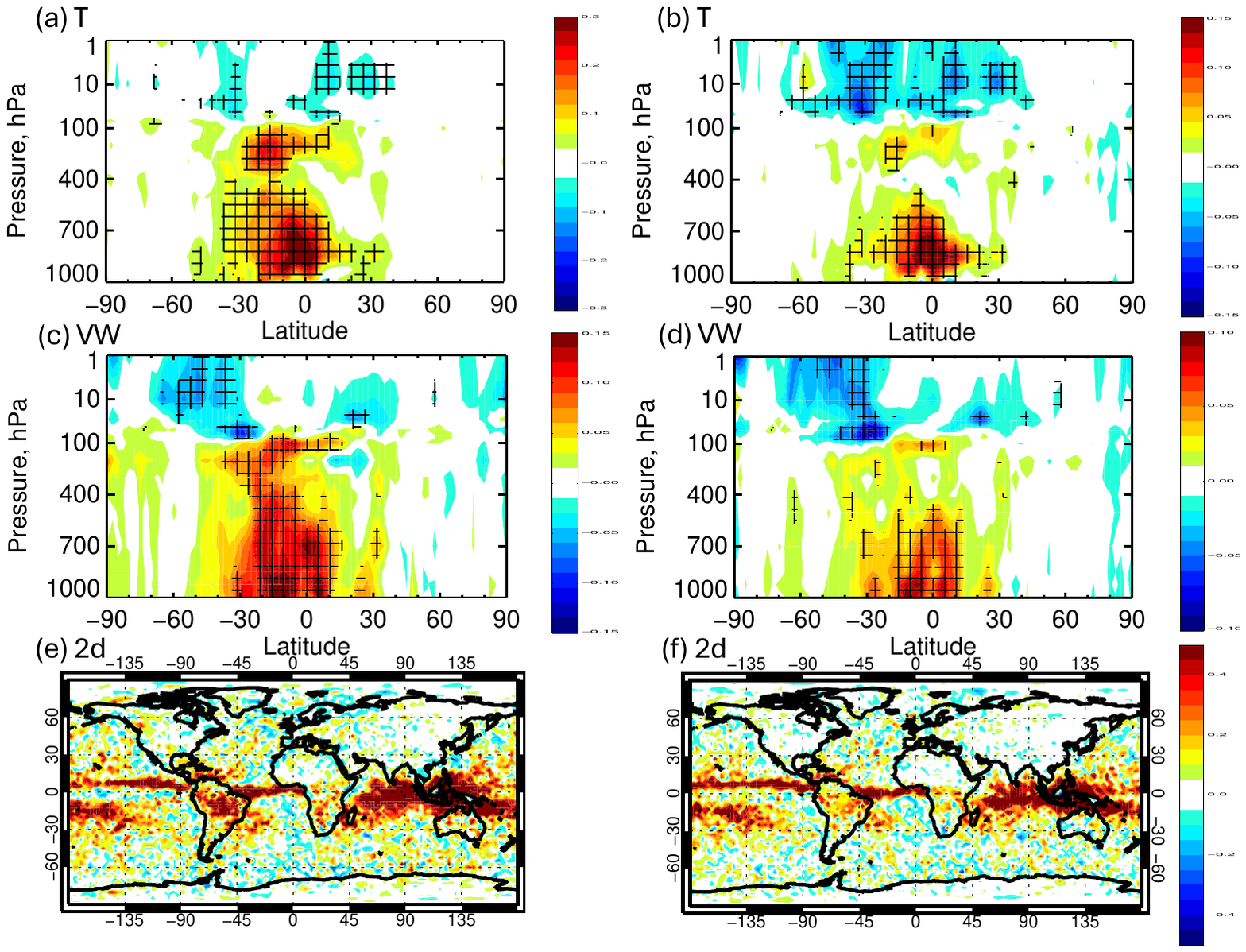}
\caption{Impact of doubling the entrainment rate in the deep convection parametrisation in forecasts without nudging (left) and with spectral nudging (right). Vertical cross-sections of (a,b) temperature and (c,d) vector wind component RMSE differences of an experiment which uses double entrainment when nudging is not active (a,c) and when nudging is active (b,d). Note the different contour intervals for (a) and (b) and for (c) and (d).  Horizontal cross section of (e,f)
dew-point temperature RMSE difference of an experiment which uses double entrainment when nudging is not active (e) and when nudging is active (f). Red colours indicate degradation (increase of RMSE) against the control experiment without double entrainment, as used
in operational ECMWF forecasts. Hatched areas indicate point of statistical significance. Scores are computed against ECMWF analysis for forecasts initialised at 00 UTC between 1 and 31 December 2025. }
\label{fig:fig12}
\end{figure}

\begin{figure}
\centering
\includegraphics[width=\textwidth]{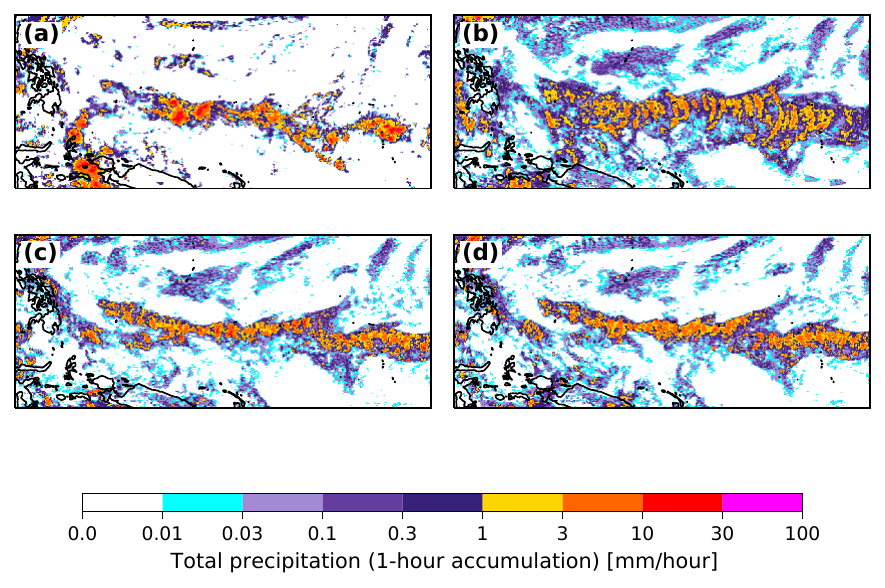}
\caption{Total precipitation accumulated over 1 hour between 23:00–00:00 UTC on 2 January 2025 for (a) IMERG observations and the three forecast experiments initialized at 00:00 UTC on 1 January 2025 using: (b) standard convection, (c) doubling entrainment rate, and (d) doubling entrainment rate combined with spectral nudging.}
\label{fig:fig13}
\end{figure}

Figure~\ref{fig:fig12} shows that the impact of the entrainment change remains clearly detectable when spectral nudging is applied. When evaluated consistently against an appropriate control configuration — that is, comparing nudged experiments with nudged controls and non-nudged experiments with non-nudged controls — the deterioration in large-scale skill associated with increased entrainment is present in both configurations, although the magnitude of the degradation in upper-level temperature and wind fields is reduced by approximately a factor of two when nudging is active.  This demonstrates that spectral nudging does not simply override internally generated large-scale tendencies arising from changes in model physics, even when those changes project onto the large-scale flow. For surface fields, such as 2-m dew point temperature, doubling the entrainment has a comparable degradation whether nudging is present or not (cf. panels e to f in Figure~\ref{fig:fig12}) which suggests that model changes affecting surface parameters remain detectable irrespective of whether nudging is applied. In addition, we find that nudging preserves the improvement in the representation of mesoscale convective systems achieved by doubling the entrainment rate of the convective mass flux, relative to the control forecast which produces spurious coupled convective gravity waves (Figure~\ref{fig:fig13}).

The persistence of the signal can be understood from the nature of spectral nudging itself. Nudging acts as a relaxation toward the large-scale state rather than as a direct prescription, allowing the model dynamics and physics to evolve while being weakly constrained at large scales. In addition, the nudging in the present configuration is applied only to selected dynamical variables and scales, while moisture and diabatic processes evolve without direct constraint from the nudging term. Changes to convective entrainment therefore modify humidity distributions and diabatic heating, which continue to influence the circulation and can project onto large scales despite the presence of nudging. As a result, the hybrid system retains sensitivity to physical parametrization changes in this example.

At the same time, this result should not be interpreted as universally applicable. The interaction between spectral nudging and model physics is likely to depend on the nature of the parametrization change, the scales on which it acts, and the magnitude of the mean-state perturbation it introduces. Small deviations may be partially damped by spectral nudging, whereas in the present case the imposed change produces a sufficiently large signal to remain clearly detectable. Modifications that primarily affect large-scale tendencies, or that strongly interact with the nudged variables, may in some cases be more difficult to diagnose in a nudged configuration. In practice, however, physics development is routinely evaluated in non-nudged configurations as part of standard cycle-evaluation procedures, and the introduction of spectral nudging does not fundamentally alter this workflow. Consequently, physics development in hybrid systems should continue to be evaluated alongside non-nudged configurations, following established cycle-evaluation practices.

From a research-to-operations perspective, these results indicate that spectral nudging can be introduced without fundamentally compromising physics-based model development, provided that comparisons are performed consistently within the nudged and free running framework. Hybrid configurations can therefore coexist with ongoing model development while preserving the ability to identify both beneficial and detrimental physics changes.

\section{Summary, Discussions and Conclusions}\label{sec:conclusion}
This study examined the use of spectral nudging to combine the ECMWF Integrated Forecasting System (IFS) with a custom machine-learned AIFS-Single model, with the objective of improving large-scale forecast skill while preserving the dynamical and physical behaviour of the NWP model at smaller scales. The central question addressed here is whether hybridization can retain the large-scale skill advantages of machine-learning forecasts without degrading small-scale variability or extreme-event representation.

Consistent with earlier results obtained by \cite{husain2024leveraging,su2026online} and \cite{polichtchouk2026hybrid}, spectral nudging of the IFS toward AIFS-Single forecasts leads to substantial improvements in large-scale forecast skill, with gains of up to approximately 1.5 days in the tropics and 12–18 hours in the extra-tropics. Crucially, these improvements are achieved without systematic damping of small-scale variability, as evidenced by kinetic energy spectra and case-study analysis. The hybrid configuration also exhibits a factor-of-two reduction in forecast bust frequency, indicating an increased predictive skill for uncertain events in this configuration. Forecast busts can arise from a range of dynamical mechanisms, including deficiencies in diabatic processes and errors in the interaction between synoptic-scale disturbances and the large-scale flow. It remains an open question whether the hybrid configuration is more effective in mitigating certain classes of forecast busts than others. Consistent with these results, the representation of high-impact near-surface weather is maintained or improved: temperature extremes show consistent error reductions across most regimes, while wind extremes exhibit modest improvements and precipitation extremes show a more mixed response, with small gains in orographically complex regions and neutral to slightly degraded performance in winter conditions.

Tropical cyclone track forecasts benefit from the improved large-scale steering flow, with reductions in track error corresponding to approximately 12–18 hours of additional forecast lead time. At the same time, intensity evolution remains comparable to that of the high-resolution physics-based model and more physically consistent than in pure machine-learning forecasts, which tend to under-predict storm intensity. Whether similar improvements extend to extra-tropical cyclone track prediction remains an open question and will be examined in future work.

From a user perspective, the hybrid configuration remains an IFS-based forecast, in which the same dynamical core, physical parametrizations, and diagnostic framework are retained. Spectral nudging introduces an additional large-scale relaxation tendency that guides the evolution of the large-scale flow while allowing mesoscale structure to evolve primarily within the NWP model. Because nudging acts only on the large scales, the hybrid model benefits from improved large-scale evolution without inheriting potential mesoscale artifacts associated with machine-learned forecasts. This separation of scales also has implications for model development, as it allows improvements in large-scale predictive skill to be obtained from machine-learned guidance while enabling physics development to focus on improving small-scale processes and physical realism, rather than compensating for large-scale mean errors. The approach is also straightforward to implement and affects only the forecast model, leaving the data assimilation system unchanged, and does not require retuning when the resolution of the underlying NWP model is changed.   Extension of the spectral nudging framework to ensemble forecasting, in which perturbed members are nudged toward probabilistic AIFS-ENS forecasts \citep{lang2024aifsB}, is examined in a companion study \citep{polichtchouk2026hybrid}. Overall, these results demonstrate that spectral nudging provides a practical pathway for combining machine-learning and physics-based forecasting systems, enabling substantial improvements in large-scale skill while preserving physically consistent small-scale behaviour.

\section*{Acknowledgements}  We thank Michail Diamantakis and Peter Dueben  for providing useful comments on the manuscript and Christian Lessig and Thomas Haiden for useful discussions. 

\section*{Conflict of interest statement}
The authors declare no conflict of interest.

\section*{Data availability statement}
The hy-IFS forecast experiments used in this study are publicly available at 10.21957/z7ks-tc34; 10.21957/aswx-qj80; and 10.21957/0xng-et91. The operational IFS analysis and forecast data are available at https://www.
ecmwf.int/en/forecasts/datasets/open-data under ECMWF’s open data
policy.

Model IFS code developed at ECMWF are the intellectual
property of ECMWF and its member states, and therefore is not publicly available. Access to a reduced
version of the IFS code may be obtained from ECMWF
under an OpenIFS licence (see http://www.ecmwf.int/en
/research/projects/openifs for further information).

The code for AIFS is available at https://github.com/ecmwf/
anemoi-core.

\bibliographystyle{wileyqj}
\bibliography{References}

\end{document}